# Spyhoppers & Stargazers

Michael J. West, Maria Mitchell Observatory
Published in *Sound* magazine, summer 2013, with references added

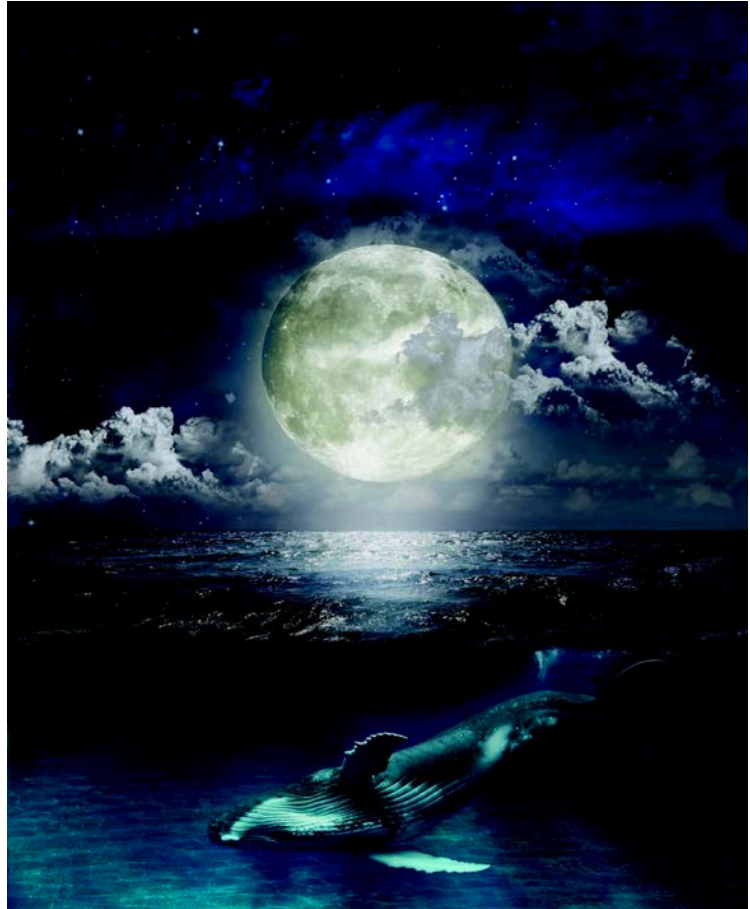

Digital Illustration: Arlene O'Reilly

In *Moby Dick*, Herman Melville wondered how – or what – whales see with eyes on opposite sides of their heads [1].

"It is plain that he can never see an object which is exactly ahead… Is his brain so much more comprehensive, combining and subtle than man's that he can at the same moment of time attentively examine two distinct prospects, one on one side of him, and the other in an exactly opposite direction?" he asked.

It's a good question. But if Melville were alive today he might have pondered something perhaps even more intriguing: **Can whales see the stars?**

Two years ago, a team led by Professor Travis Horton of the University of Canterbury in New Zealand published the most detailed study ever of the migration patterns of whales [2]. Using satellites, they tracked the movements of South Atlantic humpback whales over eight years. To their surprise, the researchers found that the whales followed almost perfectly straight paths across thousands of miles of open sea, often deviating by less than one degree. Ocean currents, storms, and varying seafloor depths - nothing seemed to knock the whales off course.

But how can humpbacks follow such straight trajectories with no landmarks to guide their way across the vast featureless seascape?

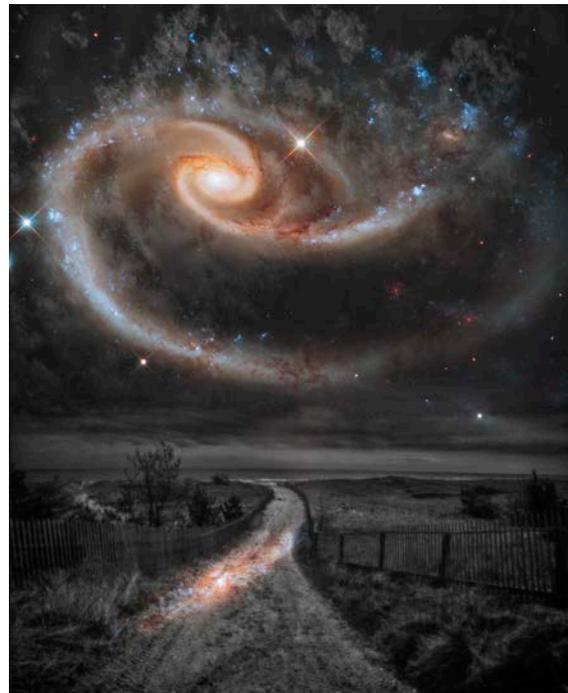

*Image Michael West / NASA / STScI*

Scientists have known for decades that migrating animals use a variety of sensory cues to orient themselves, including our planet's magnetic field and the sun's position in the sky. Yet the precision of the whales' routes seems difficult to explain with those mechanisms alone. Horton and his research team concluded, "It seems unlikely that individual magnetic and solar orientation cues can, in isolation, explain the extreme navigational precision achieved by humpback whales," speculating that "alternative mechanisms of migratory orientation" might be at work [2].

An exciting – but still unproven - possibility is that whales use the stars to chart their oceanic voyages.

It's not as farfetched as you might think. Migrating birds are known to use the stars as compasses for navigation. In a pioneering study in the 1950s, German ornithologist Franz Sauer found that European warblers [3], who migrate alone for thousands of miles mainly at night, orient themselves by the positions of the stars. Sauer performed a series of experiments in which birds were placed inside a planetarium. When the stars were visible the warblers inevitably oriented themselves in preferred geographical directions. But when simulated clouds hid the stars, the birds became completely disoriented. A 1975 study by Cornell University scientist Stephen Emlen [4] found that Indigo Buntings, birds that occasionally visit

Cape Cod, Nantucket, and Martha's Vineyard, learn to recognize the north-south direction from the apparent rotation of the night sky around the North Star.

More recently, a 2013 study in the journal *Current Biology* [5] found that South African dung beetles use the luminous haze of the Milky Way to orient themselves. These nocturnal scavengers feast on piles of dung left behind by animals, each grabbing a small piece and quickly rolling it away from other thieving beetles. The quickest route to safety is a straight line away from the dung pile, which the beetles follow using the starry skies to provide illumination and a compass.

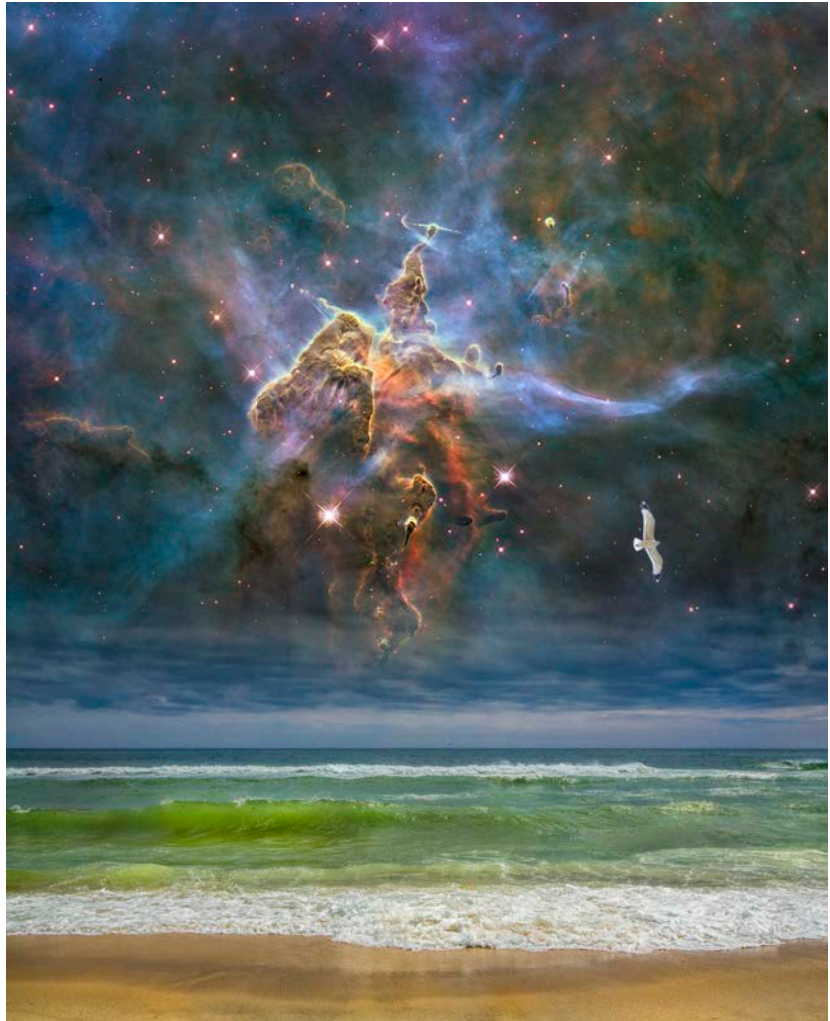
*Image Michael West / NASA / STScI*

Humans too have long used the stars to navigate. The Polynesians' legendary ability to sail across the ocean by observing the stars, winds, waves, and cloud patterns brought them to Hawaii and other Pacific islands thousands of years ago. Whaling ships leaving Nantucket or New Bedford used the stars to plot their courses. During the 19th century, slaves in southern states would escape to freedom using the Big Dipper to guide them north.

Billions of nights have passed since the first whale-like creatures returned to the ocean some 50 million years ago. Far from artificial lights created by humans, the sky over the open sea sparkles with stars. On dark moonless nights the glow of the Milky Way itself casts faint shadows across the seascape. Perhaps over time whales, like birds and people, have learned to use the stars as beacons to guide their voyages.

Of course, this presumes that whales can see the stars. Although the eyes of all mammals – whales and humans alike – share a common ancestor in the distant past, we still don't really know what whales see.

In some ways whales have it tougher than we do. As air-breathing animals, they spend time with their heads above and below the ocean's surface, which creates special challenges. For one thing, their eyes need to adapt to drastic changes in ambient light as they plunge to dark ocean depths or emerge into the bright sunlight.

And then there's the question of whether whales might be nearsighted above the ocean's surface [6]. As anyone who has ever opened his or her eyes underwater knows, the human eye is well adapted for seeing in air, but our vision becomes blurry in water. It's easy to understand why. Light rays bend whenever they pass between materials of different densities. The cornea of our eye is denser than air and acts like a lens to focus light rays on the retina, producing an image for our brains to interpret. However, the cornea and water have nearly identical densities, and consequently our eyes can't bend light enough to produce a focused image underwater.

What about whales? Having adapted to life in the ocean, is the world blurry and out of focus for them when their eyes emerge from their watery home into the air? Some evidence suggests that whales might be able to change the shape of their eyes – squinting - to see when looking through air [7]. But the ability to see in both air and water might be a compromise that leaves whales with mediocre vision in both environments, and could explain their reliance on echolocation to perceive their surroundings. It's sad to think that these majestic creatures who swim and sing beneath the stars might be unable to see them.

Whale vision has other limitations too. Although we take for granted our ability to see color, not every creature on our planet can. For humans the world is a kaleidoscope of hues, from the vibrant yellow daffodils of spring on Nantucket to the reddish glow of sunset seen from Menemsha Beach or the vivid blue of the ocean off Provincetown. A person with normal eyesight can see stars of different colors, some noticeably redder or bluer than others. Whales, on the other hand, are colorblind because their eyes lack the three kinds of cones that allow human eyes to perceive color. To a whale, the world appears in different shades of gray, like an old black-and-white movie [8].

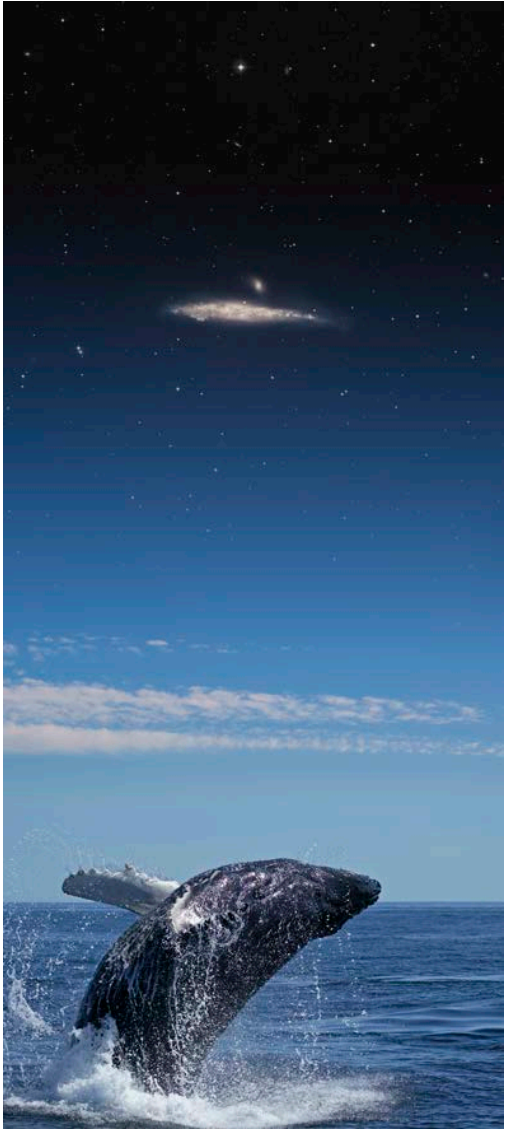

*NGC 4631 is a galaxy located about 30 million light years from Earth. Like our own Milky Way galaxy, it's a collection of billions of stars moving in a graceful gravitational dance. NGC 4631's uncanny resemblance to the gentle giants of the sea has earned it a nickname: the Whale Galaxy.*

*Image Michael West / Mathew Hull*

Perhaps the best evidence that whales might be able to see the stars comes from a behavior known as *spyhopping*. Whales, dolphins and even some sharks are observed to poke their heads out of the water to look around, sometimes for several minutes. It's the equivalent of treading water for humans. What's not clear is why they do this. Some species of whales, such as orcas, probably spyhop to look for prey. Others, like humpback whales, are often observed to spyhop when tourist boats are nearby, as if trying to figure out what all the fuss is about. Some whales spyhop with their eyes just below the water's surface, maybe for the same reason that humans wear scuba masks to see underwater; the little extra layer of air – or water in the case of whales – bends light rays enough to focus them. Some scientists, however, suggest that spyhopping is more about hearing than vision, with whales using their sensitive ears – their main sensory organ in the dark undersea world– to listen to what's going on near the ocean's surface.

If whales can see the stars are they curious or indifferent to them? There's abundant evidence that whales and other cetaceans are intelligent creatures. A recent YouTube video seen by millions shows a distressed dolphin approaching divers in an obvious plea to help untangle it from fishing line. The brains of humpback whales are known to have certain types of cells also found in humans, apes and elephants that are related to higher mental functions such as intelligence and emotions. Last year scientists even reported that a captive beluga whale began to make unusual sounds that they recorded and interpreted as his attempt to mimic human speech [9].

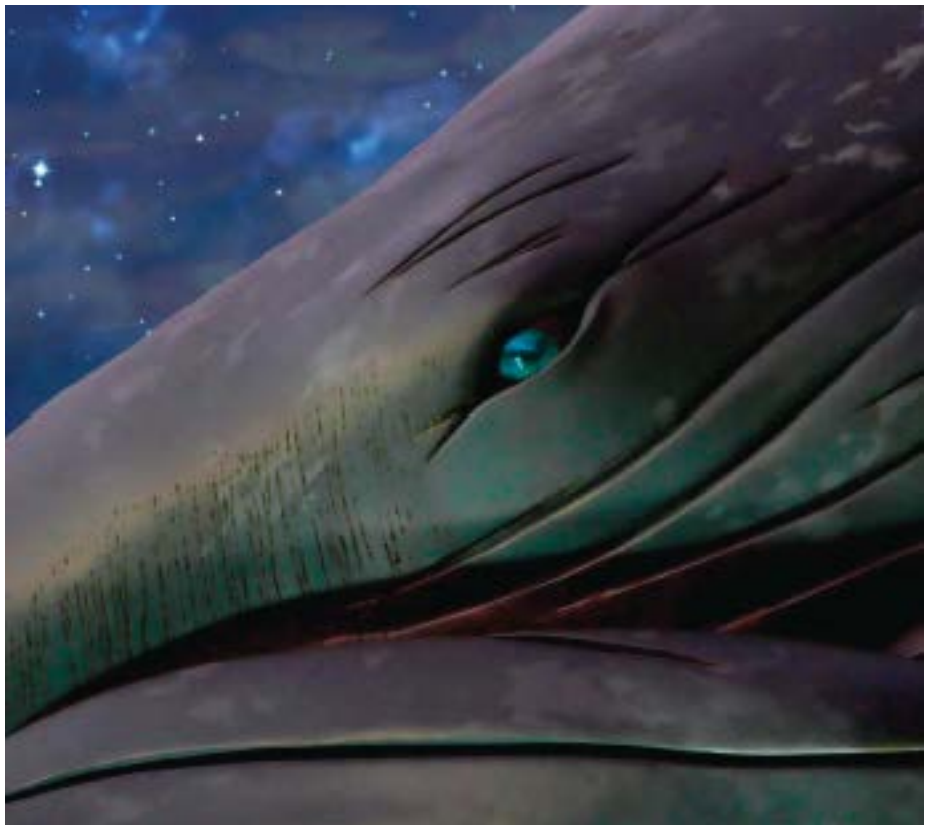
Digital Illustration: Arlene O'Reilly

However, intelligence doesn't necessarily mean that whales share our interest in astronomy or want to know more about the starry skies. We humans build telescopes because we're curious about our celestial home. Whales may be intelligent but they don't build telescopes, either because their lack of opposable thumbs makes it impossible or because they simply have no interest in learning more about the stars.

But I wonder if there might be more to it.

One of astronomy's most profound discoveries is that earth and all its creatures are made from the ashes of stars whose fires burned out billions of years ago. Perhaps that's why we humans feel compelled to explore the starry skies, as if driven by an innate yearning to know our true ancestral home. And maybe that's the real reason whales spyhop. Aquatic astronomers hoping to glimpse more than the shores of the land they left behind, answering an ancient call to raise their heads skyward, guided by only a faint, primal memory that they – like us - can't articulate or even fully comprehend. Nostalgia for the stars.